Title Page

# Analysis Of Radiation Dose Distribution Inhomogenity Effects In Gamma Knife Radiosurgery Using Geant4

**Özlem Dağlı, Dr.** Gazi University, Faculty of Medicine, Department of Brain and Neurosurgery, Beşevler/ Ankara/ Turkey, ozlem_150184@hotmail.com, +90 312 202 44 73

[*]Corresponding author

**Ayşe Güneş Tanır, Prof. Dr.** Gazi University, Faculty of Sciences, Department of Physics, 06500 Teknikokullar/ Ankara/ Turkey, gunes@gazi.edu.tr, Phone: +90 312 202 12 36,

**Gökhan Kurt, Prof. Dr.** Gazi University, Faculty of Medicine, Department of Brain and Neurosurgery, Beşevler/ Ankara/ Turkey, gkurtmd@gmail.com , +90 312 202 51 16

# Analysis Of Radiation Dose Distribution Inhomogenity Effects In Gamma Knife Radiosurgery Using Geant4

*Research Article*

Özlem DAĞLI[1*], Ayşe Güneş TANIR[2], Gökhan Kurt[1]

[1] Gazi University, Faculty of Medicine, Department of Brain and Neurosurgery, Beşevler/ Ankara/ Turkey

[2] Gazi University, Faculty of Sciences, Department of Physics, 06500 Teknikokullar/ Ankara/ Turkey

## ABSTRACT

The Monte Carlo method is widely used in the Gamma Knife dose distribution calculations. In this study, Monte-Carlo simulation with Geant4 was applied to determine Leksell Gamma Knife dose distribution for homogeneous solid water and heterogeneous brain phantoms, and primarily focused on the differences arising from the effects of inhomogeneity. The relative dose graphs were displayed for different collimator diameters. The dose values from the maximal interval regions were determined and compared. It seen that when the collimator size was 4mm, the mean maximum relative dose point was at 47mm for water phantom and at 44mm for the brain. The difference found was about 6.38%. The differences among the results determined for water and brain were ranged from 2 to 17%. As a result, the inhomogeneity effect should be considered in dose calculation in Gamma Knife planning system.

**Keywords**: Radiosurgery, Gamma Knife, γ-Rays, Inhomogeneity, Brain

**INTRODUCTION**

Gamma Knife was developed in 1968 by the neurosurgeon, Lars Leksell [1], and the physicist, Börje Larsson, using 179 Co-60 sources. The Gamma Knife (LGK) system is used together with the treatment planning system, called the Leksell Gamma Plan (LGP). In LGK, the dose calculation algorithm does not consider the scatter dose contributions or the inhomogeneity effect resulting from the skull and air cavities [2–4].

The increasing importance of radiation-related applications in our lives has increased the importance of measuring the amount of radiation absorbed by living organisms or any material exposed to radiation. There are numerous national and international regulations and recommendations governing homogeneous mediums to ensure the accuracy of radiation dosimetry [5]. The human body and brain; however, have different density components, such as bone and airspaces. Many researchers have widely used the Monte Carlo simulation for inhomogeneity problem [6–8]. The results from them show discrepancies: some of them reported that the measurements are in agreement with simulation results, while some are not. Solberg et al.[3] have pointed out a remarkable disagreement between Monte Carlo results and Cheung et al.[9] have reported that there are discrepancies among them up to 25%.

Radiosurgery involves directing target bundles of beams from several different angles, resulting in rapid dose reduction in normal tissues outside the target, while high doses are achieved in the confluence of the rays. The size of the target volume is of great importance in terms of prevention of normal tissue complications, if the tumor is to be irradiated with complete accuracy at each session. Gamma Knife radiosurgery in advanced imaging and planning techniques can be directed into the small space inside the skull by using a very thin gamma ray beam.

Geant4 Monte Carlo simulation software features a large library of physics, which includes tools capable of simulating precisely the interaction of the particles with the target. At the basis of Geant4 software, a category scheme is required to enable the designed detector, to structure the capability of identifying both geometric and physical phenomena, and to read the results. Starting from the bottom of the scheme and moving upwards, the necessary structures and procedures for the simulation are carried out, resulting in the creation of a subordinate structure of the simulation operation in a certain sequence to establish order.

In this study, dosimetry calculations were made for the application of Gamma Knife device using solid water and brain phantoms. The design of the skull and device source was modeled with the Geant4 program, and the relative dose graphs were determined. The dose distribution was simulated for 4,8,14 and 18mm collimators and the results were compared with the results of water and brain phantoms.

**MATERIALS AND METHODS**

In the study, a Gamma Knife device, solid water phantom, brain phantom, electrometer, ion chamber and the Geant4 computer program were used at Gazi University Medical Faculty Hospital. The Gamma Knife device used was the 4C model (Elekta) with 201 sources.

Phantoms are devices that are equivalent to human tissue and which are used to examine dose distributions in tissue. A large percentage of the human body is made up of water and the atomic number Z of muscles and soft tissues is almost equivalent to water. Because of this, the basic dose distributions are usually made in the solid water phantom, as they are reproducible and are very close to the radiation absorption-scattering properties of muscles and soft tissues and they can be measured in three dimensions. Ideally, mass density, number of electrons per mass and effective atomic number must be equal to water, in order to ensure that the material to be used is tissue or water equivalent. In this study, an 160mm diameter solid water and brain phantom were used. It has entrance holes for positioning the detector inside. The contents of the water and brain phantoms were taken from [10, 11].

Composition of Water, Liquid(ICRP): Density( $g/cm^3$ ) = 1.00000E+00, Mean Exicitation Energy(eV)=75.000000

**COMPOSITION:**

| Atomic number | Fraction by weight |
|---|---|
| 1-Hydrogen | 0.111894 |
| 8-Oxygen | 0.888106 |

**Table1.** Composition of Water

Composition of Brain(ICRP): Density( $g/cm^3$ ) = 1.03000E+00, Mean Exicitation Energy (eV)=73.300000.

**COMPOSITION:**

| Atomic number | Fraction by weight |
|---|---|
| 1-Hydrogen | 0.110667 |
| 6-Carbon | 0.125420 |
| 7-Nitrogen | 0.013280 |
| 8-Oxygen | 0.737723 |
| 11-Sodium | 0.001840 |
| 12-Magnesium | 0.000150 |
| 15-Phosphorus | 0.003540 |
| 16-Sulfur | 0.001770 |
| 17-Chlorine | 0.002360 |
| 19-Potassium | 0.003100 |
| 20-Calcium | 0.000090 |
| 26-Iron | 0.000050 |
| 30-Zinc | 0.000010 |

**Table 2.** Composition of Brain

The dose values in the gamma knife treatment protocol were compared with the solid water phantom by using ion chamber. SAD= 40 cm (distance between source and target) Electrometer switched on when ion chamber and cable were not connected. Zero setting of the electrometer by waiting for 10 minutes. Then the ion chamber was placed in the center of the phantom. Not taking the first measurement, 10 measurements were taken for one minute.

|  | 1. | 2. | 3. | 4. | 5. | 6. | 7. | 8. | 9. | 10. | ort |
|---|---|---|---|---|---|---|---|---|---|---|---|
| read value (Mu) (nC) | 10,09 | 10,09 | 10,085 | 10,085 | 10,09 | 10,09 | 10,095 | 10,095 | 10,09 | 10,09 | 10,09 |

**Table 3.** Dose values read from the electrometer

$D_w = M_{u,o} \times N_{d,w} \times P_{t,p}$ = 10,09 x 255,936118 x 1,109693 = 2,865 Gy/min

| Gamma Knife Data | |
|---|---|
| Calibration dose | 3.385 Gy/min at 2014-09-18 |
| Days since calibration at approval date | 420 |
| Treatment dose rate (2015-11-12) | 2.909 Gy/min |
| Effective output factors (4, 8, 16) | 0.814 0.900 1.000 |
| Ring 1 output factors (4, 8, 16) | 0.812, 0.934, 0.961 |
| Ring 2 output factors (4, 8, 16) | 0.823, 0.919, 1.000 |
| Ring 3 output factors (4, 8, 16) | 0.795, 0.874, 0.981 |
| Ring 4 output factors (4, 8, 16) | 0.726, 0.782, 0.914 |
| Ring 5 output factors (4, 8, 16) | 0.664, 0.708, 0.847 |

**Fig 3**. Treatment protocol from Gamma knife planning computer

Treatment dose rate 2,909 Gy/min acording to treatment protocol from Gamma knife planning computer (Fig 3). Treatment dose rate read from electrometer 2,865 Gy/min. The difference was 1,5%.

All experimental procedures were conducted by ten year experienced clinical physician.

**Geant4**

In the Geant4 application, to get started, first, the geometrical elements were prepared (i.e. the materials, volumes, and locations to be simulated); the relevant physics elements were identified (i.e. particles, physical processes, models, and production threshold energy: mechanism of formation of primary particles); the prepared geometry and particle traces were displayed; and user interface (UI) commands were added for the user. During the simulation, the necessary information needs to be collected. Geant4 used in this study tracked particle down to zero kinetic energy or until the particle leaves the simulation geometry. This approach is more accurate than discarding electrons with energy below the threshold because the latter could significantly increase uncertainties for dose calculation in voxels, mainly, in low density media [12]. The $4\times10^6$ event was considered and it takes about 46.9h. The calculations were made for four different collimators (4, 8, 14, 18 mm) and it was repeated 4 times for each collimator.

Taking advantage of the documentation of the present system, a new modular design has been made for the simulation application of the Gamma Knife device. The aim here was to identify the gamma beam profile and to reveal the most convenient and useful structure for obtaining the beam profile using the existing beam-collimator materials.

In the Geant4 simulation program, four basic steps were taken into consideration when preparing the simulation software of the Gamma Knife system: Definition of general structure, a formation of physical geometry, definition of physical phenomena, and simulation process and calculations. 3D models of the Gamma Knife design were prepared, and front graphic models for simulation were created in a computer environment. Figure 1 shows the 3D design of the source. Three basic geometric structures, World, Target, and Tracker, are defined by codes in Geant4 software. According to this definition, the World, a room with an indoor air of 400×400 cm$^2$, represents the model's location and the model of the environment in which the experiment will take place. Target is a 90×90×90 mm voxel detector plate (Fig. 2, 3).

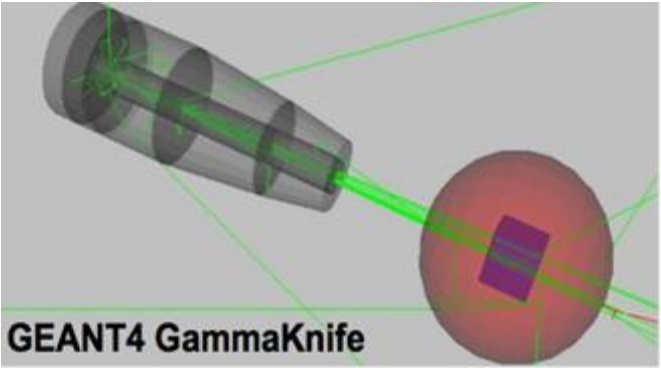

**Fig.1** 3D source design with Geant4

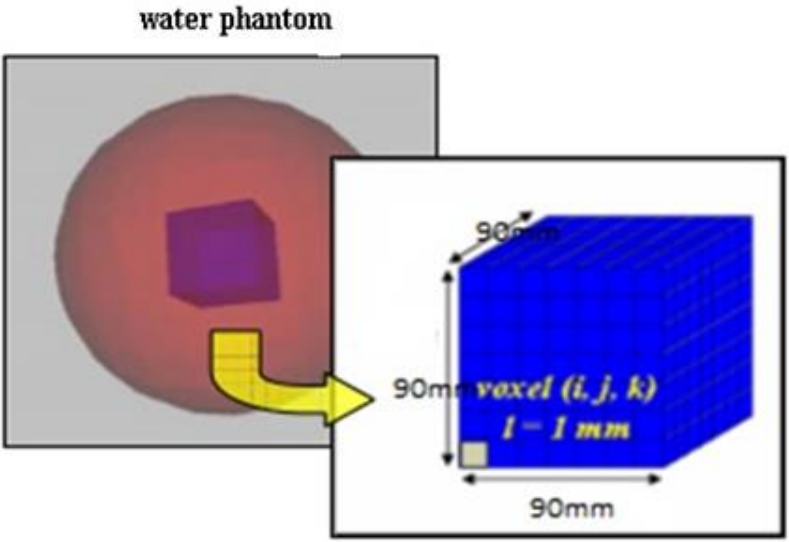

**Fig.2** Voxel display in solid water phantom

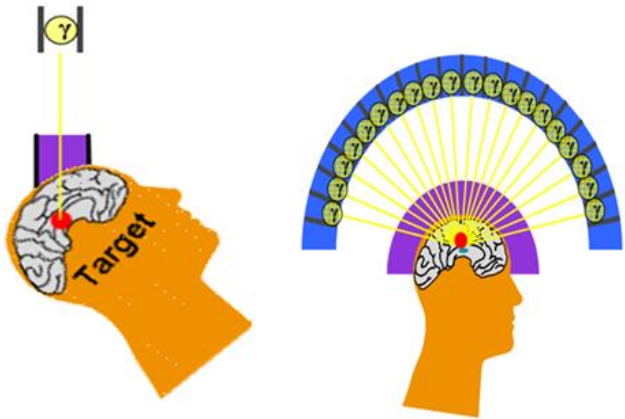

**Fig 3.** Focus of distribution of resources and focus as a single source

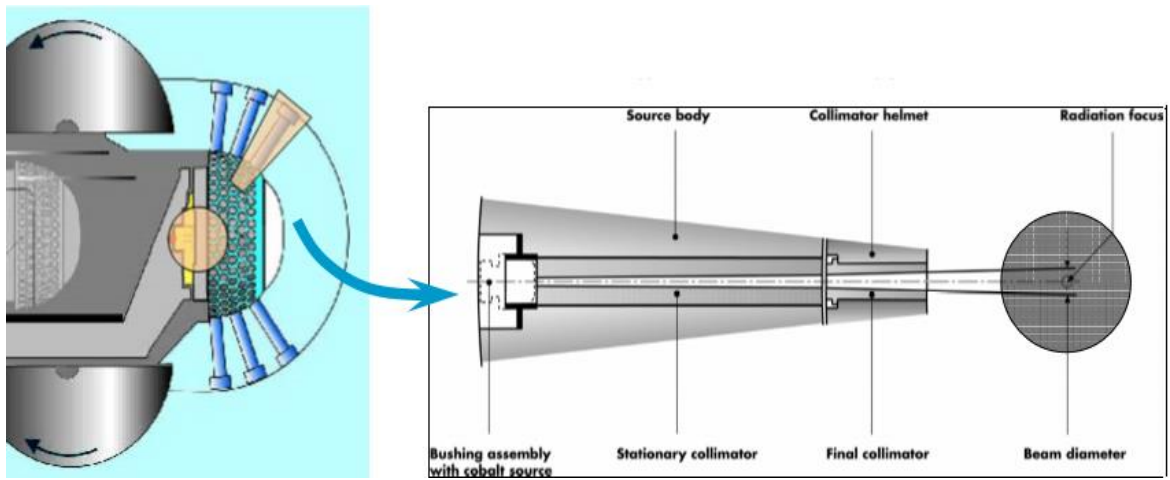

**Fig 4.** Location of a single source and phantom on the device

In the final step of the simulation study, the modeling processes and physical calculations were as follows: The Gamma Knife device consisted of a semi-spherical iron-plated unit with 201 Co-60 sources. Gamma rays came from different directions and focused on the target. Calculations were made for four different collimators: 4, 8, 14, and 18 mm (Fig. 5). The stored energy was calculated at the end of the simulation using voxels divided into small cubes of 90×90×90 mm with a scoring mesh.

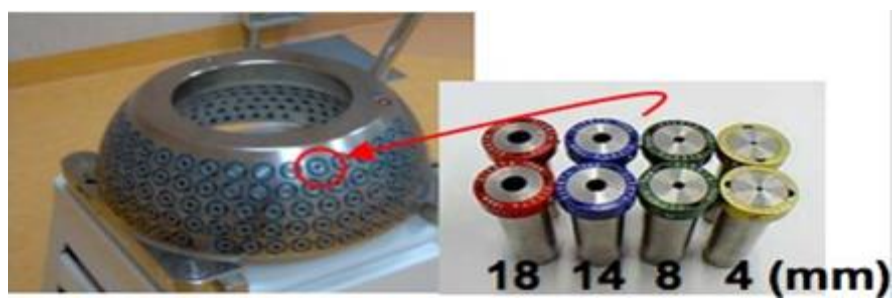

**Fig. 5** Display of collimators

When modeling, a spherical water phantom of 160mm in diameter was used instead of the human brain for dosimetry measurements. The patient source distance was 401mm, and two gamma rays of 1.17 MeV and 1.33 MeV were released from Co-60 decay.

**RESULTS**

The Geant4 simulation program designed the three-dimensional model of the Gamma Knife. Since modeling in the program was based on both solid water (homogeny) and human brain phantom (heterogenic), dose differences can be attributed to inhomogeneity effects. Human tissue equivalent phantom was used in examining the dose distributions. The dose values using the solid water phantom were compared with the values using human brain phantom.

Figures 6, 7, 8 and 9 display the normalized (relative) dose profiles of the simulations using collimators with 4mm, 8mm, 14mm, and 18mm diameter.

In examining the graphs obtained with Geant4, the different diameters of the collimators showed the following results: when the 4mm collimator (Fig.6) was used, the mean maximum relative dose (MMRD) was at 47 mm for solid water and at 44 mm for brain phantom. The difference between them is about 6.4%, which possibly resulted from the inhomogeneity difference between brain and water phantoms. When the 8mm collimator (Fig.7) was used, the MMRD was at 46mm for solid water and at 45mm for the brain; using the 14mm collimator (Fig.8), the MMRD was at 47mm for water phantom and at 46mm for the brain. The difference between these two is about 2%; and lastly, when the 18mm collimator (Fig.9) was used, the MMRD was at 46mm for water and at 38mm for the brain, with the difference between these two is about 17%.

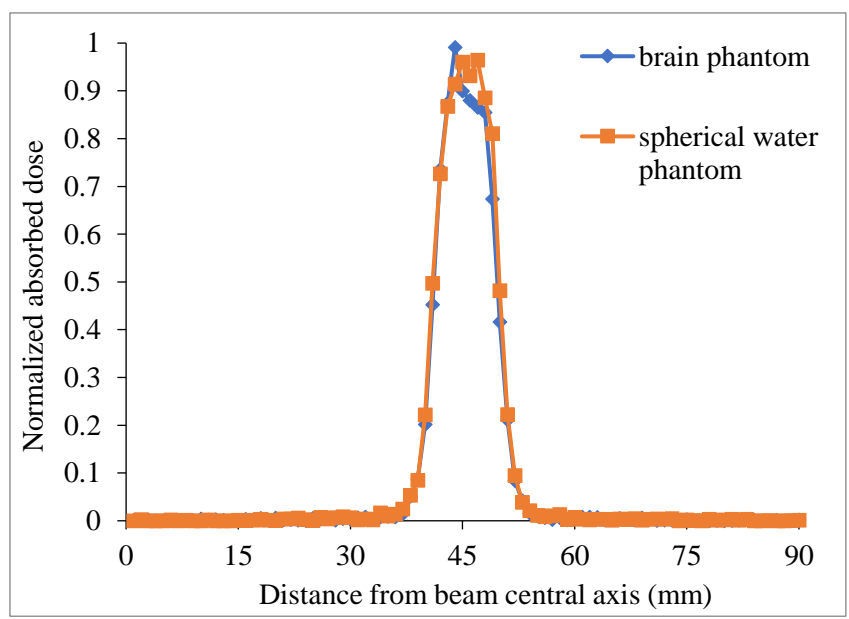

**Fig.6** Comparison of relative dose graphs for solid water and brain phantom along the x-axis for the collimator with of 4mm diameter.

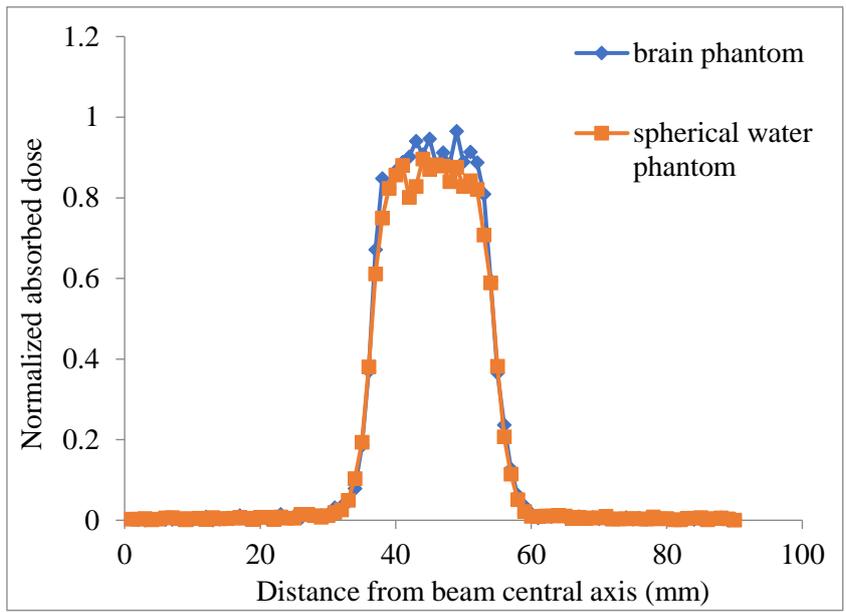

**Fig.7** Comparison of relative dose graphs for solid water and brain phantom along the x-axis for the collimator with of 8mm diameter.

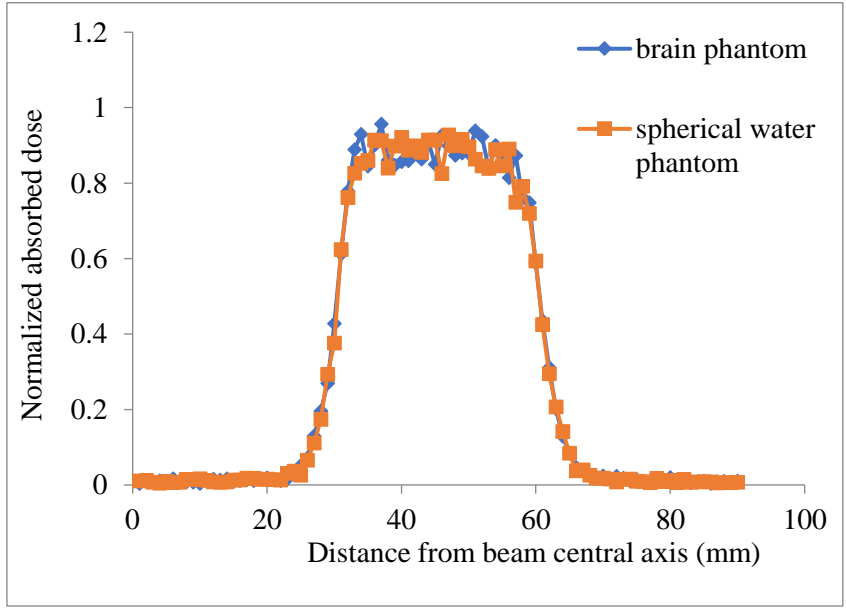

**Fig.8** Comparison of relative dose graphs for solid water and brain phantom along the x- axis for the collimator with of 14mm diameter.

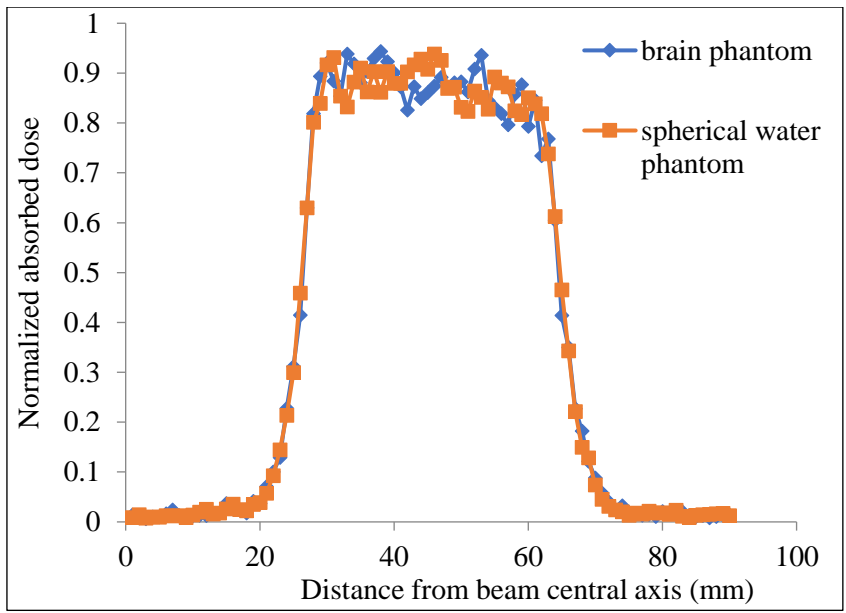

**Fig.9** Comparison of relative dose graphs for solid water and brain phantom along the x- axis for the collimator with of 18mm diameter

The similar comparisons were made considering the relative absorbed dose at the same point. When the collimator opening was 4mm, the mean relative dose of water phantom was 0.9644±0.0735 at distance 47mm, while the mean dose absorbed for the brain was 0.8656±0.0928 at the 47mm. The difference found was about 10.2%.

When the collimator opening was 8mm, the mean relative dose absorbed by the water phantom was 0.8790±0.0786 at 47mm, while the dose absorbed for the brain was 0.9123±0.0606 at 47mm. The difference found was about 3.8%.

When the collimator opening was 14mm, the value for water was 0.9273±0.0706 at 47mm, while the dose absorbed for the brain was 0.8991±0.066 at the same point. The difference found was about 3%.

When the collimator opening was 18mm, the value for water phantom was 0.9253±0.0637, while the absorbed dose for the brain was 0.8915±0.0654. The difference found was about 3.65%.

As seen from Figures 6, 7, 8 and 9 the statistical distribution is also increasing as the collimator diameter increases.

**DISCUSSION**

Monte Carlo simulations have been applied to radiotherapy treatment planning for some time and are capable of accurately predicting various dosimetry parameters, as well as doses in regions where electronic equilibrium is lacking. The necessity of securing more accurate dose calculations using algorithms such as Monte Carlo is now recognized by the medical physics community [13–15]. Previous Monte Carlo simulations of the Gamma Plan$^R$ output show no differences in homogeneous phantoms [13]. However, more recent studies have shown substantial differences between commonly used treatment planning algorithms and Monte Carlo results when heterogeneous patient anatomy is applied.

According to the studies by Cheung et al., [9, 16–18], "A set of Monte Carlo data computed for verifying Gamma Plan$^R$ output showed no differences with data obtained using a homogeneous phantom." However, an analysis of published data by Cheung et al., [9] shows differences up to 25% between Leksell Gamma plan dose calculations and Monte Carlo results. Regulations by AAPM indicated that relative differences as small as 5% can be accepted as reliable treatment [19].

In this study, dose measurements were made with ion chamber and electrometer to determine the accuracy of the solid water phantom. When the result is compared with the we get from Gamma knife treatment planning system less than the acceptable range (< 3%) prescribed by the stereotactic radiosurgery. Than the study made with the geant4 simulation program, the three-dimensional model of the Gamma Knife device was designed. It can be said more clearly that the differences arise from the effects of inhomogenity, since modeling in the program was done by considering both the water phantom and the human brain phantom. Human tissue equivalent brain phantoms are among those used because the real human brain cannot be used in the examination of dose distributions. Solid water phantoms have homogeneous structures. But not every part of the human brain has the same density. Brain consists of structures sinus cavities, head bones, hard connective tissues such as dura mater, brain parenchyma, vascular structures. So it is not homogeneous.

The results of this study do not find the difference as big as Cheung's findings. However, as the collimator diameter increases, the difference increases significantly. That is, as the target volume grows, the effect of inhomogeneity is critical.

Xiaowei and Chunxiang [6] reported the simulation of dose distribution irradiation by LGK unit. They calculated the dose distributions using the EGS4 Monte Carlo code. They showed that "the calculated results are in good agreement with those using semiconductor"[6].

Al-Dweri et al. [7], studied the Monte Carlo simulation using the code PENELOPE to test a simplified model of the source channel geometry of LGK [7]. Al-Dweri et al. [8], also applied Monte Carlo simulation with PENELOPE (version 2003) to calculate LGK dose distribution for heterogeneous phantoms. They reported that "The dose distribution determined for heterogeneous phantoms including the bone-/or air-tissue interfaces show non-negligible differences with respect to those calculated for homogeneous one". The conclusions of the results of them support the results of this study.

## CONCLUSION

Accurate determination of the radiation dose given to the patient is of critical importance. The results of this study show that the differences between the determination of the distance to be given the dose using Geant4 for solid water phantom and brain are remarkable. These differences, which can be attributed to inhomogeneity effects, are larger than the acceptable range (< 3%) prescribed by the stereotactic radiosurgery according to AAPM Report 54, AAPM Report 142 . Therefore, the inhomogeneity effect should consider for Gamma Knife treatment planning.

## ACKNOWLEDGMENT

The authors would like to thank Dr. Gazi Erkan Bostancı from Ankara University for the Monte Carlo simulation with Geant-4. This research did not receive any specific grant from funding agencies in public or not-for-profit sectors.

**Conflict of interest:** Authors have not received research grants from any company. Author Dr. Özlem Dağlı declares that she has no conflict of interest. Author Dr. Ayşe Güneş Tanır declares that she has no conflict of interest. Author Dr. Gökhan Kurt declares that he has no conflict of interest

**Ethical approval**: This article does not contain any studies with human participants or animals performed by any of the authors.